\documentclass{aastex631}

\shorttitle{6 Contact Binaries}
\shortauthors{Wadhwa et al.}
\graphicspath{{./}{figures/}}
\usepackage{hyperref}

\begin{document}

\title{A Study of Six Extreme Low Mass Ratio Contact Binary Systems}

\author[0000-0002-7011-7541]{Surjit S. Wadhwa}
\affiliation{School of Science, Western Sydney University,\\ Locked Bag 1797, Penrith, NSW 2751, Australia.\\}
\author[0000-0002-8036-4132]{Bojan Arbutina}
\affiliation{Department of Astronomy, Faculty of Mathematics, University of Belgrade, Studentski trg 16, 11000 Belgrade, Serbia.\\}
\author[0000-0001-8535-7807]{Jelena Petrovi\'c}
\affiliation{Astronomical Observatory, Volgina 7, 11060 Belgrade, Serbia\\}
\author[0000-0002-4990-9288]{Miroslav D. Filipovi\'c}
\affiliation{School of Science, Western Sydney University,\\ Locked Bag 1797, Penrith, NSW 2751, Australia.\\}
\author[0000-0001-9677-1499]{Ain Y. De Horta}
\affiliation{School of Science, Western Sydney University,\\ Locked Bag 1797, Penrith, NSW 2751, Australia.\\}
\author[0000-0002-9931-5162]{Nick F. H.  Tothill}
\affiliation{School of Science, Western Sydney University,\\ Locked Bag 1797, Penrith, NSW 2751, Australia.\\}

\author[ 0000-0001-9392-6678]{Gojko Djura\v sevi\'c}
\affiliation{Astronomical Observatory, Volgina 7, 11060 Belgrade, Serbia\\}



\begin{abstract}

Multi-band (B, V and R) photometric and spectroscopic observations of six poorly studied contact binaries carried out at the Western Sydney University and Las Cumbres Observatory were analysed using a recent version of the Wilson-Devenney code. All six were found to be of extreme low mass ratio ranging from 0.073 to 0.149. All are of F spectral class with the mass of the primary component ranging from 1.05$M_{\sun}$ to 1.48$M_{\sun}$. None show light curve features of enhanced choromospheric activity (O'Connel Effect) however five of the six do have significant ultraviolet excess indicating presence of increased magnetic and chromospheric activity. Period analysis based on available survey data suggests two systems have a slowly increasing period suggesting mass transfer from the secondary to the primary, two have a slow declining period with likely mass transfer from primary to the secondary while one shows a steady period and one undergoing transition from a declining to increasing period suggesting possible mass transfer reversal. We also compare light curve solutions against theoretical markers of orbital stability and show that three of six systems have mass ratios within the theoretical instability limit and maybe regarded as potential merger candidates. 
\end{abstract}



\section{Introduction} \label{sec:intro}

Investigation of extreme low mass ratio contact binaries has recently seen heightened interest with view to identifying potential merger (red nova) progenitors \citep{2021MNRAS.501..229W, 2021MNRAS.502.2879G, 2022MNRAS.512.1244C, 2023MNRAS.519.5760L}. It has been known for some time that merger events and orbital instability in contact binaries is most likely when the mass ratio of the components ($q=M_2/M_1$) is below some critical value \citep{1995ApJ...438..887R, 2007MNRAS.377.1635A, 2009MNRAS.394..501A}. We have recently introduced methods to aid in the rapid identification of potential low mass ratio contact binary systems from survey photometry data \citep{2022JApA...43...94W} in addition to a theoretical framework linking the mass of the primary component and geometric elements determined through light curve analysis to orbital instability \citep{2021MNRAS.501..229W}. 

We have previously reported analysis of fifteen extreme low mass ratio contact binaries with features of orbital instability \citep{2022RAA....22j5009W, 2023arXiv230615190W}. This study reports photometric and spectroscopic observations of six extreme low mass ratio poorly studied contact binary systems selected from the All Sky Automated Survey for SuperNovae (ASAS-SN) \citep{2014ApJ...788...48S, 2020MNRAS.491...13J}. The systems were selected for observations based on the techniques described in \cite{2022JApA...43...94W} as being likely of low mass ratio and potentially unstable. Identification details for the systems are summarised  in \hyperref[Table 1]{Table 1}. In addition to light curve analysis we show that at least 5 systems exhibit features of chromospheric activity without photospheric evidence for star spots.

\begin{table}
\label{Table 1}
 \caption{Identifications, abbreviations, check and comparison stars for 6 studied} systems
 %\centering
 %\scriptsize
   \centering

   \begin{tabular}{|l|l|l|l|}
    \hline
        \hfil Name &\hfil Abbreviation &\hfil Comparison Star &\hfil Check Star \\ 
        \hline
        \hfil ASAS J054049-5527.8 & \hfil A0540 &\hfil 2MASS 05403192-5527279 & \hfil 2MASS 05405378-5526306 \\
        
         \hfil ASAS J084220-0303.4	 &\hfil A0842   &\hfil TYC 4867-806-11  &\hfil TYC 4867-463-1\\ 
          \hfil ASAS J103737-3709.5 &\hfil A1037   &\hfil TYC 7197-1470-1  &\hfil TYC 7197-1596-1 \\ 
          \hfil ASAS J104422-0711.2 &\hfil A1044  &\hfil TYC 4919-253-1 &\hfil 2MASS 10441404-0710597\\ 
           \hfil V565 Dra &\hfil V565 Dra  & \hfil TYC 3897-742-1 &\hfil 2MASS 17383576+5710441 \\
          \hfil ASAS J200304-0256.0 &\hfil A2003 &\hfil TYC 5164-275-1 &\hfil 2MASS 10032574-0257207 \\ 
            \hline
           
    \end{tabular}
    \end{table}

\section {Photometric and Spectroscopic Observations}

A1044 was observed over 5 nights in April 2020 with the Western Sydney University (WSU) 0.6m telescope equipped with a cooled SBIG 8300 CCD camera and standard Johnson $BVR$ filters.  All other systems were imaged using the 0.4m telescopes from the Las Cumbres Observatory (LCO) network. The LCO network telescopes acquire images using the SBIG STL-6303 CCD camera and Bessel $V,B$ and Sloan $r'$ filters. Images were acquired in $V$ and $R/r'$ bands for all systems except A0842 which was only observed in $V$ band due to technical difficulties. To document the $B-V$ magnitude we also acquired between 40 and 50 images during eclipses in the $B$ band. All images of A1044 were calibrated using multiple dark, flat and bias frames. The LCO network has an automated pipeline which provided calibrated images for all other systems. Differential photometry for each system was performed with the AstroImageJ \citep{2017AJ....153...77C} package using the comparison and check stars noted in \hyperref[Table 1]{Table 1}. The comparison star magnitudes were adopted from the American Association of Variable Star Observers (AAVSO) Photometric All-Sky Survey \citep{2015AAS...22533616H}. The AstroImageJ package estimates photometric errors and we excluded all observations where the estimated error was greater than 0.01 magnitude. Details of observations such as dates, observation numbers, exposure times along with light curve characteristics such as amplitude, maximum brightness, and $B-V$ colour are collated in \hyperref[Table 2]{Table 2}. 

\begin{table} [!ht]
\label{Table 2}

 %\centering
 %\scriptsize
\caption{Details of observations, spectral class and light curve parameters. $(B-V)_o$ is distance and extinction corrected estimate.}
\centering
   \begin{tabular}{|l|l|l|l|l|l|l|l|l|}
    \hline
        \hfil Name &\hfil Date of Obs &\hfil Obs . ($V,R$) &\hfil Exp. Times ($V,R$) &\hfil Max Bright. ($V$) &\hfil Ampl. ($V$) &\hfil $B-V$&\hfil$(B-V)_o$&\hfil Sp. Type \\
        \hline
        \hfil A0540 &\hfil 11/22 - 01/23 &\hfil 310, 340  &\hfil 45s,40s&\hfil 12.52&\hfil 0.35&\hfil 0.62&\hfil 0.58&\hfil F9 \\ 
        \hfil A0842   &\hfil 02/23 - 02/23 &\hfil 375, - &\hfil 45s,-&\hfil 11.51&\hfil 0.30&\hfil 0.51&\hfil 0.50&\hfil F8 \\ 
         \hfil A1037   &\hfil 02/23 - 05/23  &\hfil 420, 360&\hfil 40s,35s&\hfil 10.79&\hfil 0.34&\hfil 0.58&\hfil 0.55&\hfil F9  \\ 
          \hfil A1044 &\hfil 04/20 - 05/20  &\hfil 415, 395 &\hfil 40s,33s&\hfil 11.72&\hfil 0.23&\hfil 0.15&\hfil 0.12&\hfil F0* \\ 
          \hfil V565 Dra &\hfil 06/22 - 05/23 &\hfil 460, 345&\hfil 45s,40s&\hfil 11.26&\hfil 0.32&\hfil 0.63&\hfil 0.60&\hfil F7* \\ 
           \hfil A2003  & \hfil 08/22 - 08/22 &\hfil 480, 550&\hfil 40s,35s&\hfil 10.51&\hfil 0.41&\hfil 0.68&\hfil 0.53&\hfil F7 \\
            \hline      
    \end{tabular}

    $^*$Spectral classification from LAMOST survey.
    \end{table}

\begin{figure}[!ht]
    \label{Figure 1}
	\includegraphics[width=\textwidth]{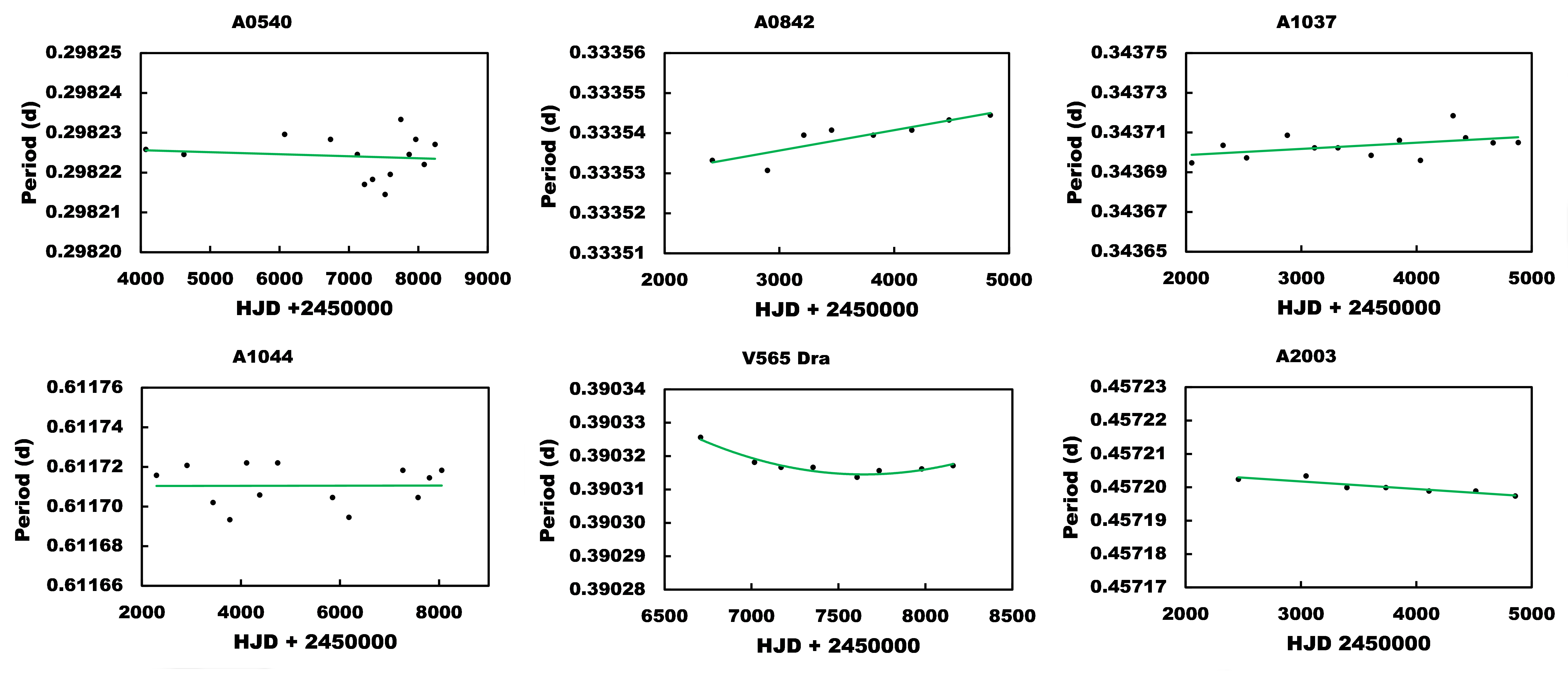}
    \caption{Period trend based on survey photometry data. The green line represents the best fit.}
    \end{figure}

Assessment of period variation, especially when small, requires high cadence long term (over many decades) observations. Given the lack of suitable historical observations no meaningful Observed-Computed $(O-C)$ analysis could be performed for any of the systems. Instead we use the technique of employing periodic orthogonal polynomials and an analysis of variance statistic (a quality of fit marker) to fit multiple overlapping subsets, each of approximately 100 - 150 observations, of $V$ or $g'$ band survey photometry data for each system to estimate any significant period variations. The methodology is described in detail by \citet{1996ApJ...460L.107S} and was used by \citet{2011A&A...528A.114T} to demonstrate the exponential decay in the period of the only confirmed contact binary merger system V1309 Sco. We find that two of our systems (A540 and A2003) have a linear trend towards a reducing period, two systems (A0842 and A1037) have a linear trend of a rising period, V565 Dra has a shallow parabolic trend indicating a shift from falling to rising period while A1044 appears to have a relatively steady period. If one considers the transfer of mass as the only contributor to period change a rising period suggests mass transfer from the secondary to the primary and visa versa for a falling period. The period trends are summarised in \hyperref[Table 3]{Table 3} and illustrated in \hyperref[Figure 1]{Figure 1}.

    \begin{table} [!ht]
    \label{Table 3}
\caption{Updated orbital elements and period trend.}
 %\centering
 %\scriptsize
   \centering

   \begin{tabular}{|l|l|l|l|}
    \hline
        \hfil Name &\hfil Epoch (HJD) &\hfil Period (d)&\hfil Period Trend (d/yr) \\
        \hline
        \hfil A0540 &\hfil $2459948.146810\pm0.000250$ &\hfil$ 0.2982187\pm0.0000050$&\hfil $-1.87\times10^{-7}$ \\
        
        \hfil A0842   &\hfil $2459984.366616\pm0.000309$ &\hfil $0.3335395\pm0.0000025$&\hfil $1.86\times10^{-6}$  \\ 
         \hfil A1037   &\hfil $2459986.138027\pm0.000212$  &\hfil $0.3434028\pm0.0000015$&\hfil $1.15\times10^{-6}$ \\ 
          \hfil A1044 &\hfil$2458944.604343\pm0.000052$  &\hfil $0.6117118\pm0.0000010$&\hfil Steady\\ 
          \hfil V565 Dra &\hfil $2459752.674348\pm0.000322$ &\hfil$0.3903187\pm0.0000025$&\hfil Parabolic \\ 
           \hfil A2003  & \hfil $2459811.109575\pm0.000230$ &\hfil$0.4571959\pm0.0000030$&\hfil $-8.27\times10^{-7}$ \\
           \hline
           
    \end{tabular}
    
    \end{table}

Successful light curve analysis of contact binary systems without radial velocity data is only possible if complete eclipses are present \citep{2005Ap&SS.296..221T}. During such analysis the temperature of the primary ($T_1$) is usually fixed as the shape of contact binary light curves is dependent almost exclusively on the geometric parameters such as the mass ratio, inclination and degree of contact. The light curve shape places a constrain on the component temperature ratio ($T_2/T_1$) but not on the absolute value of the component temperatures \citep{1993PASP..105.1433R, 2001AJ....122.1007R}. Notwithstanding the above, varied methods are used in assigning the temperature of the primary ($T_1$) with colour based estimations being employed most often. Colour calibrated estimates, although widely used, have been shown to be cumbersome. Recently in an analysis of four contact binaries \citet{2022PASA...39...57H} found temperature variations between $B-V$ and $J-K$ colour calibrations in excess of $500$K for 2 stars and $250$K and $150$K for the other two. \citet{2023RAA....23c5012M} reported variation in excess of $400$K between spectra and space based survey colour databases. The VizeR database records a range in excess of $1000$K for four of the systems reported here and many hundreds for the other two. Spectral classification of stars possibly represents an accurate and standard method and more recently many investigators  \citep[see e.g.][]{2023PASP..135e4201L, 2023PASP..135d4201G, 2022PASJ...74.1421C, 2022MNRAS.517.1928G} have adopted low resolution spectral class calibrations (where available) as an alternative to assign the usually fixed value for $T_1$.

\begin{figure}[!ht]
    \label{Figure 2}
	\includegraphics[width=\textwidth]{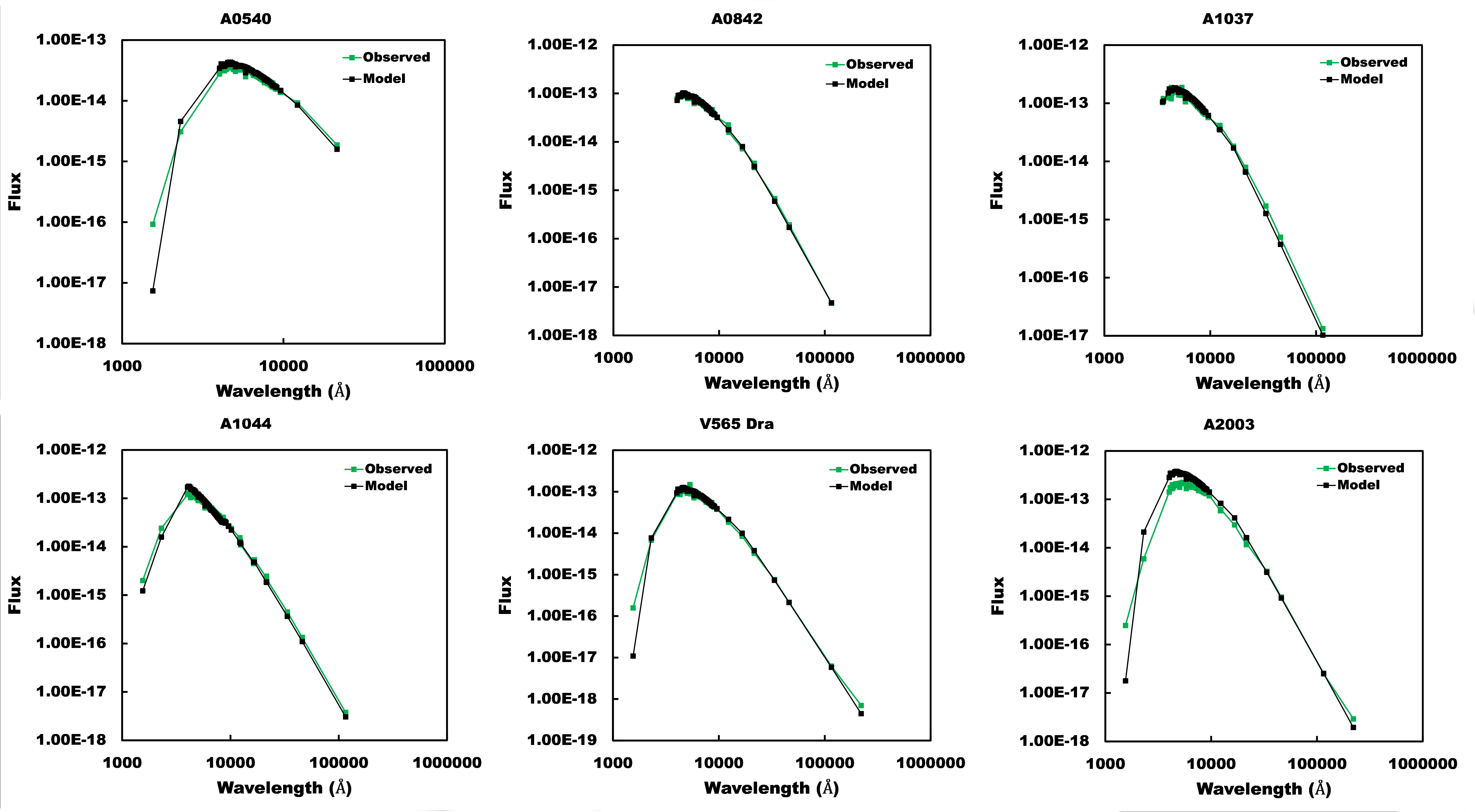}
    \caption{Observed and modelled SEDs for all six systems. The observed photometry is indicated in green and the fitted model in black. The flux on the vertical axes is in erg/cm$^2$/s/\AA. The wavelength is in Angstroms (\AA). Both axes are in log scale.}
    \end{figure}

\begin{figure}[!ht]
    \label{Figure 3}
	\includegraphics[width=\textwidth]{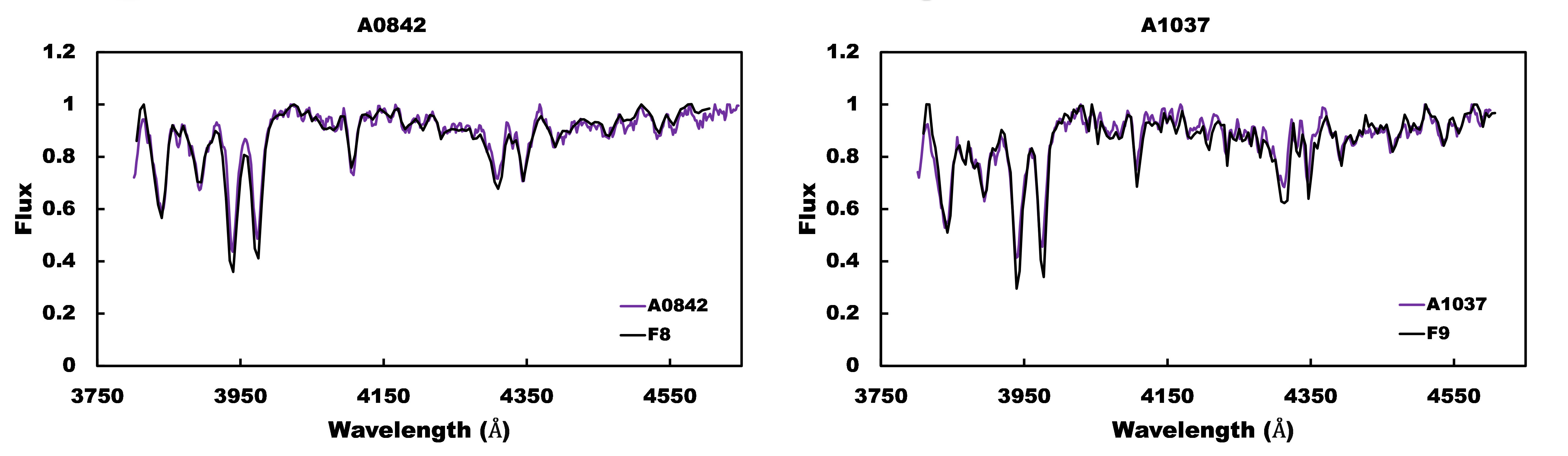}
    \caption{Observed (low resolution) and matched library spectra for A0842 and A1037.}
    \end{figure}

One mechanism to overcome the wide variations that can result from various templates and colour calibrations is through the investigation of the Spectral Energy Distribution (SED) constructed using photometric data from various bands collectively. \citet{2007ApJS..169..328R} and \citet{2008A&A...492..277B} performed comparison of the SEDs constructed from survey photometry with synthetic theoretical spectra and the modelled value for the effective temperature compared favourably to the theoretical spectral value. We compared the effective temperature of the systems (and hence temperature of the primary) determined through SED calibration against those estimated through spectral class for each system. Firstly, using the methodology described in \citet{2008A&A...492..277B} we constructed a photometry data set (SED) in different bands for each system from publicly available survey data. The constructed SEDs were then fitted to theoretical models which incorporated Kurucz atmospheres  using ${\chi}^2$ minimisation as described in \citet{2008A&A...492..277B} to determine the effective temperature. The SEDs and fitted model are illustrated in \hyperref[Figure 2]{Figure 2}. Two (A1044 and V565 Dra) of our six systems were observed with the The Large Sky Area Multi-Object Fiber Spectroscopic Telescope (LAMOST) \citep{2018yCat.5153....0L} with the reported spectral classes F0 and F7, respectively. For the other four systems we used the 2m telescopes from the LCO network equipped with the low resolution FLOYDS spectrograph to acquire spectra that were compared to standard main sequence star spectra from \citet{1984ApJS...56..257J, 1998PASP..110..863P} to determine the spectral class for each system as recorded in \hyperref[Table 2]{Table 2}. Selected FLOYDS spectra and matched library spectra are shown in \hyperref[Figure 3]{Figure 3}. We used the April 2022 update of \citet{2013ApJS..208....9P} calibration tables of spectral class and temperature for main sequence stars to determine the spectral based temperature of the primary component. The VizieR range, SED and spectral class effective temperatures are summarised in \hyperref[Table 4]{Table 4}. From \hyperref[Table 4]{Table 4} we see that in all cases except A2003, where the variation between SED and spectral effective temperature varies by more than $500$K, there is good agreement between a collective photometric approach and spectral class for estimating the effective temperature. The findings are very similar to \citet{2022ApJ...927...12P} who also found close agreement between spectral class effective temperature and SED modelled values. The difference in A2003 most likely, although not for certain, relates to the relative high extinction for the system with an estimated distance corrected extinction of 0.48 magnitudes. Overall we consider that the recent trend towards the use of low resolution spectra and spectral classification to assign a fixed value to the temperature of the primary to be valid and we have used this method for light curve analysis of the systems presented in this study.

     \begin{table}
     \label{Table 4}
\caption{Effective temperature (K) range as reported in the VizieR database, SED modelling and spectral class interpolation.}
 %\centering
 %\scriptsize
   \centering

   \begin{tabular}{|l|l|l|l|}
    \hline
        \hfil Name &\hfil VizieR Range &\hfil SED&\hfil Spec. Class \\
        \hline
        \hfil A0540 &\hfil 5818 - 6874 &\hfil6000&\hfil 6050 \\
        
        \hfil A0842   &\hfil 5959 - 6740 &\hfil 6250&\hfil 6180  \\ 
         \hfil A1037   &\hfil 5305 - 6338  &\hfil 6000&\hfil 6050 \\ 
          \hfil A1044 &\hfil 6632 - 7793  &\hfil7250&\hfil 7200 \\ 
          \hfil V565 Dra &\hfil 5921 - 7044 &\hfil6250&\hfil 6280 \\ 
           \hfil A2003  & \hfil5006 - 5773 &\hfil5750&\hfil 6280 \\
           \hline
           
    \end{tabular}
    
    \end{table}

\section{Light Curve Analysis}
We confirm that all systems show total eclipses and as such determination of the mass ratio from photometry alone would be possible. As there is no significant asymmetry in the maximum brightness only unspotted solutions were modelled. We used the Wilson-Devinney (WD) code (2013 version) which incorporates Kurucz atmospheres to model simultaneous $V$ and $R$ band light curve solutions \citep{2021NewA...8601565N, 1998ApJ...508..308K, 1990ApJ...356..613W}. Since the effective temperature of the primary is below 7500K for each system, the gravity darkening coefficients were set equal $g_1 = g_2 = 0.32$, and bolometric albedoes as $A_1 = A_2 = 0.5$. we used the logarithmic limb darkening coefficients from \citet{1993AJ....106.2096V} as advocated by \citet{2015IBVS.6134....1N}. 

\begin{figure}[!ht]
    \label{Figure 4}
	\includegraphics[width=\textwidth]{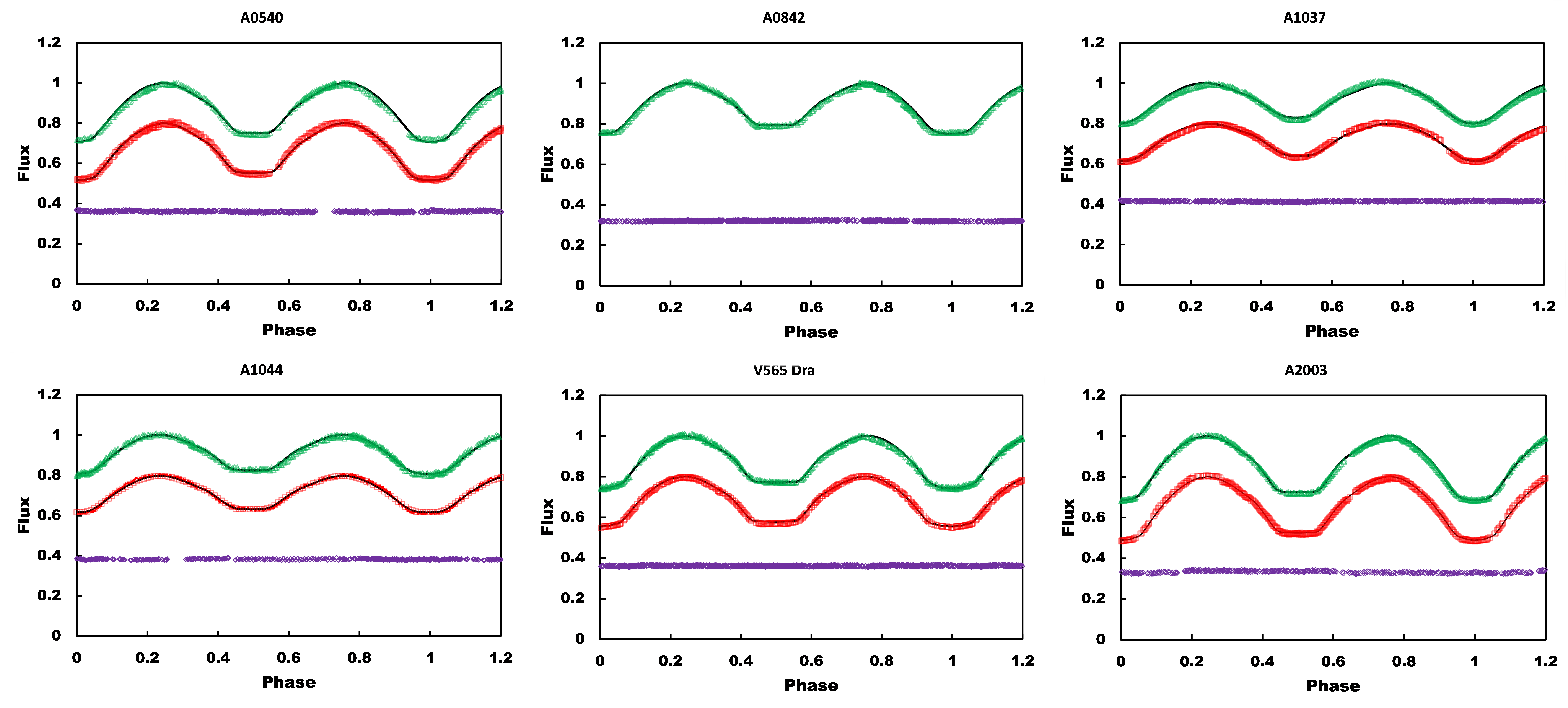}
    \caption{The WD model (black line) and observed V band (open green triangle), R band (open red square) light curves for the six reported systems. The open purple diamonds represents the check star. The flux has been arbitrarily shifted vertically for clarity}
    \end{figure}

We searched for the mass ratio ($q$) for the systems using the grid method first described by \citep{1982A&A...107..197R}. Systematic search was made for a range of fixed values for the mass ratios from 0.05 to 15. A coarse search was performed up to $q=1$ in increments of 0.1 and in increments of 0.2 up to $q=15$. The search was then refined in increments of 0.01 near the best solution. During the search procedure the temperature of the secondary component ($T_2$), the surface potential ($\Omega$) (i.e. fillout $f$),  orbital inclination ($i$) and the dimensionless luminosity of the primary ($L_1$) were all treated as adjustable parameters. For each mass ratio, iterations were executed until the reported standard deviations were higher than the suggested adjustment for all parameters. To obtain the full solution the mass ratio was also made an adjustable parameter during the last iteration and the suggested standard deviations for each parameter was recorded as the potential error. Summary of the light curve solutions is presented in \hyperref[Table 5]{Table 5}. Observed and WD fitted light curves are illustrated in \hyperref[Figure 4]{Figure 4}.

    \begin{table}
    \label{Table 5}
\caption{Light curve solution and other absolute parameters for six investigated contact binary systems.}
 %\centering
 %\scriptsize
   \centering

   \begin{tabular}{|l|l|l|l|l|l|l|}
    \hline
        \hfil  &\hfil A0540 &\hfil A0842 &\hfil A1037&\hfil A1044 &\hfil V565 Dra &\hfil A2003\\
        \hline
        \hfil $T_1$ (K) (Fixed) &\hfil 6050 &\hfil 6180  &\hfil 6050&\hfil 7200&\hfil 6280&\hfil 6280 \\
        
        \hfil$T_2$  (K)  &\hfil $5884\pm16$ &\hfil $5997\pm27$ &\hfil $5741\pm34$&\hfil $7190\pm27$&\hfil $6259\pm11$&\hfil $6266\pm10$\\ 
         \hfil Incl. $(^\circ)$   &\hfil $82.3\pm0.9$  &\hfil $77.5\pm1.4$&\hfil $68.1\pm0.8$&\hfil $72.7\pm0.5$&\hfil $90.0^{+0.0}_{-0.5}$&\hfil $82.5\pm0.5$\\ 
          \hfil $q$ &\hfil $0.14\pm0.001$  &\hfil $0.100\pm0.006$ &\hfil $0.090\pm0.003$&\hfil $0.073\pm0.003$&\hfil$0.108\pm0.002$&\hfil $0.149\pm0.002$\\ 
           \hfil $q_{inst}$ ($f$=0)  &\hfil $0.083\pm0.006$  &\hfil $0.084\pm0.003$ &\hfil $0.074\pm0.003$&\hfil $0.048\pm0.001$&\hfil$0.080\pm0.013$&\hfil $0.093\pm0.005$\\ 
           \hfil $q_{inst}$ ($f$=1) &\hfil $0.094\pm0.008$  &\hfil $0.096\pm0.004$ &\hfil $0.086\pm0.004$&\hfil $0.053\pm0.001$&\hfil$0.091\pm0.017$&\hfil $0.108\pm0.006$\\ 
          \hfil Fillout (\%) &\hfil $69\pm2$ &\hfil $83\pm4$&\hfil $57\pm4$&\hfil $83\pm3$&\hfil $71\pm2$&\hfil $82\pm2$ \\ 
           \hfil $r_1$ (mean)  & \hfil 0.578 &\hfil 0.605&\hfil 0.604&\hfil 0.625&\hfil 0.596&\hfil 0.578\\
           \hfil $r_2$ (mean)  & \hfil 0.259 &\hfil 0.239&\hfil 0.219&\hfil 0.216&\hfil 0.238&\hfil 0.271\\
           \hfil $M_1/M_{\sun}$&\hfil $1.13\pm0.05$ &\hfil $1.12\pm0.03$&\hfil $1.18\pm0.02$&\hfil $1.48\pm0.05$&\hfil $1.15\pm0.11$&\hfil $1.05\pm0.03$ \\ 
           \hfil $M_2/M_{\sun}$&\hfil $0.16\pm0.02$ &\hfil $0.11\pm0.02$&\hfil $0.11\pm0.01$&\hfil $0.11\pm0.01$&\hfil $0.12\pm0.0.03$&\hfil $0.16\pm0.02$ \\ 
           \hfil $M_{V1}$&\hfil $3.76\pm0.20$ &\hfil $4.16\pm0.10$&\hfil $3.97\pm0.10$&\hfil $2.43\pm0.21$&\hfil $3.77\pm0.48$&\hfil $4.41\pm0.15$\\
            \hfil $A/R_{\sun}$&\hfil $2.04\pm0.02$ &\hfil $2.17\pm0.01$&\hfil $2.25\pm0.02$&\hfil $3.53\pm0.03$&\hfil $2.44\pm0.09$&\hfil $2.65\pm0.01$\\
            \hfil $R_1/R_{\sun}$&\hfil $1.18\pm0.02$ &\hfil $1.31\pm0.01$&\hfil $1.36\pm0.02$&\hfil $2.21\pm0.03$&\hfil $1.45\pm0.05$&\hfil $1.53\pm0.01$\\
            \hfil $R_2/R_{\sun}$&\hfil $0.53\pm0.02$ &\hfil $0.52\pm0.01$&\hfil $0.49\pm0.02$&\hfil $0.76\pm0.02$&\hfil $0.58\pm0.03$&\hfil $0.72\pm0.01$\\
            \hfil $\Delta\rho$ & \hfil -0.54 &\hfil -0.43&\hfil -0.60&\hfil -0.15&\hfil -0.37&\hfil -0.18\\
            \hfil UV Excess & \hfil -1.75 &\hfil -1.53&\hfil --&\hfil -4.17&\hfil -2.28&\hfil -1.43\\
            \hline
           
    \end{tabular}
    \end{table}

\section{Absolute Parameters and Orbital Stability}
\subsection{Absolute Parameters}
Full investigation of astrophysical phenomenon such as orbital stability and chromospheric activity requires knowledge of the absolute physical parameters especially the mass of the primary. Without high resolution spectroscopic observations one is reliant on indirect methods to estimate the mass of the primary component. In this study we use the mean of a distance based estimate and a colour calibration based estimate of the mass of the primary. It is accepted that the primary component of contact binaries follow zero age main sequence profile \citep{2013MNRAS.430.2029Y}. For our colour based estimation we used the  2MASS $J-H$ magnitudes \citep{2006AJ....131.1163S} for each system and the calibration tables of \citet{2013ApJS..208....9P} (April 2022 update) for low mass ($0.6M_{\sun} < M_1 < 1.6M_{\sun}$) stars to interpolate the mass of the primary component. 

The distance based estimation was interpolated from the absolute magnitude of the primary component corrected for extinction. The absolute magnitude of the primary component was determined as follows: As all systems have total eclipses and are of low mass ratio the secondary eclipse apparent magnitude represents the apparent magnitude of the primary. We obtained the absolute magnitude of the primary ($M_{V1}$) using the GAIA EDR 3 \citep{2022A&A...658A..91A} distance and the line of sight extinction corrected for distance ($E(B-V)_d$) as described in \citet{2023arXiv230615190W}. The absolute magnitude was obtained using the standard distance module. The observed $B-V$ was also corrected for extinction $(B-V)_o$ as:
\begin{equation}
    (B-V)_o = (B-V) - E(B-V)_d .
\end{equation}
The extinction corrected $(B-V)_o$ values are  summarised in \hyperref[Table 2]{Table 2} while the estimated absolute magnitudes are summarised in \hyperref[Table 5]{Table 5} along with other absolute parameters.

The distance based mass of the primary was obtained from interpolation of the absolute magnitude and mass of main sequence stars from the calibration tables of \citet{2013ApJS..208....9P} (April 2022 update) for low mass ($0.6M_{\sun} < M_1 < 1.6M_{\sun}$) stars. The distance based estimate resulted in the largest error and this was adopted as the error for the mass estimation. All other errors were propagated from this estimation. The mass of the secondary ($M_2$) was determined from the mass ratio and Kepler's third law used to derive the current separation ($A$) between the components. The light curve solution provides an estimate of the fractional radii of the components ($r_{1,2}$) for three orientations. The geometric mean of these was used to estimate the absolute radii ($R_{1,2}$) of the components by applying ${R}_1 = r_1 A$ and ${R}_2 = r_2 A$ as per  \citet{2005JKAS...38...43A}. All the absolute parameters are summarised in \hyperref[Table 5]{Table 5}.

It been well established that the secondary components have larger radii than main sequence stars of similar mass. \citet{2022JApA...43...94W} also reported that the radius of the primary may also be more than 25\% larger than corresponding main sequence stars. In addition to change in the radii of the components some researchers \citep{2013MNRAS.430.2029Y} suggest that  evolutionary mechanisms will likely lead to significant density variation between the components such that the secondary will always be denser and the the difference between density of the primary and secondary components ($\Delta\rho$) will always be less than zero \citep{2004A&A...414..317K}.

 As noted above the light curve solution provides fractional radii for each component and one can use geometric mean of these along with Equation (3) from \citep{1981ApJ...245..650M} to calculate the density difference ($\Delta\rho$). All our system show that the density of the secondary is indeed higher and that $\Delta\rho$ is negative. The results are summarised in \hyperref[Table 5]{Table 5}.

\subsection {Orbital Stability}
The merger potential of contact binary systems and their orbital stability has received significant attention recently \citep{2021MNRAS.501..229W, 2023MNRAS.519.5760L, 2022MNRAS.512.1244C}. New mathematical relations linking the instability mass ratio, mass of the primary and the degree of contact have recently been reported by \citet{2021MNRAS.501..229W} who showed that for low mass primaries ($0.6M_{\sun} < M_1 < 1.6M_{\sun}$) the instability mass ratio ($q_{inst}$) is between:
\begin{equation}
\label{eq:qinst-f1}
    q_{inst}=0.1269M_{1}^2-0.4496M_{1}+0.4403\  (f=1)
\end{equation} 
and
\begin{equation}
\label{eq:qinst-f0}
  q_{inst}=0.0772M_{1}^2-0.3003M_{1}+0.3237\  (f=0).  
\end{equation}
The above equations represent the extremes of the instability mass ratio at marginal contact ($f=0$) and full overcontact  ($f=1$).
\\
\\
We calculate the instability mass ratio range ($q_{inst}$) for each system and provide it \hyperref[Table 5]{Table 5}. Although all six systems have extreme low mass ratios only three (A0842, A1037 and V565 Dra) can be classified as being potential merger candidates with modelled mass ratios within the error range for the instability mass ratio. A0540, A1044 and A2003 all have modelled mass ratios well above the instability mass ratio range and as such must be considered likely stable. As all systems described are relatively bright and well within the reach of modest instruments regular monitoring of the potential merger candidates even by advanced amateurs should be encouraged.

\section{High Energy Indicators of Chromospheric Activity}

Contact binary systems usually have periods of less than 24 hours with synchronised rotation. Magnetic activity is thought to be high in rapidly rotating systems including contact binaries \citep{2019BlgAJ..31...97G} resulting in increased stellar magnetic wind and magnetic breaking. Increased magnetic breaking will eventually lead to loss of angular momentum from the system and potential orbital instability \citep{2004MNRAS.355.1383L}. The only significant photospheric indicator of increased magnetic activity is the presence of star spots usually manifesting as the O'Connell effect or asymmetric maxima of the light curve. The photosphere is dominated by high intensity low energy emissions which obscure lesser intensive chromospheric emissions therefore light curve analysis provides little indication of chromospheric activity. Direct measure of angular momentum loss is difficult, nevertheless,  secondary indicators of enhanced magnetic and chromospheric activity \citep{1983HiA.....6..643V, 1983MNRAS.202.1221R, 2004MNRAS.355.1383L} are potentially easier to observe. The six systems described in this report do not demonstrate photometric features of enhanced magnetic/chromospheric activity. Significant chromospheric and magnetic activity however is not excluded. Emissions at higher energy levels such as the far ultraviolet band can provide a clearer indicator of such activity. The GALEX (Galaxy Evolution Explorer) satellite surveyed the sky in both the far-ultraviolet band (FUV) centered on 1539 {\AA} and near-ultraviolet band (NUV) centered on 2316 {\AA}. Only the FUV band can be relied upon for the detection of chromospheric activity as NUV emissions may also be contaminated by photospheric emissions \citep{2010PASP..122.1303S}. 

As demonstrated by \citep{1984ApJ...279..763N, 1996AJ....111..439H} the $R_{\rm HK}^{\prime}$ index and $\log R_{\rm HK}^{\prime} \geq -4.75$ are characteristic indicators of a more active star. \citet{2010PASP..122.1303S} matched GALEX FUV magnitudes ($m_ {\rm FUV}$) to the $\log R_{\rm HK}^{\prime}$ for dwarf stars to derive the $\Delta(m_{\rm FUV-B})$ colour excess:
\begin{equation}
    \Delta(m_{\rm FUV-B}) = (m_{\rm FUV}-B) - (m_{\rm FUV}-B)_{\rm base}
\end{equation}
where
\begin{equation}
    (m_{ \rm FUV}-B)_{\rm base} = 6.73(B-V) + 7.43,
\end{equation}
They concluded that chromospherically active stars have a UV colour excess below -0.5 while those with lesser chromospheric and magnetic activity have a colour excess above -0.5.

The GALEX mission observed five of our six systems with measured FUV magnitudes. Ultraviolet colour excess was calculated for all five using Equations 4 \& 5. All five have ultraviolet colour excess well below -0.5, a value suggestive of enhanced magnetic/chromospheric activity in the absence of photospheric features. It is well known that magnetically active features such as plages \citep{2008LRSP....5....2H} may be associated with star spots, however, the reverse is not always the case -- chromospheric/magnetic activity without photospheric features is possible \citep{2017ApJ...835..158M}. \hyperref[Table 5]{Table 5} summarises the the UV colour excesses for the five systems observed by GALEX

\section{Summary and Conclusion}

The confirmation that the red nova V1309 Sco in 2008 was indeed a merger event between components of a contact binary system \citep{2011A&A...528A.114T} has significantly increased interest in the study of contact binaries. Although it has been known for some time that merger events are likely at low mass ratios \citep{1995ApJ...438..887R, 2007MNRAS.377.1635A, 2009MNRAS.394..501A} most investigators until more recently have sought to define a minimum mass ratio at which orbital stability is likely. Recent theoretical updates \citep{2021MNRAS.501..229W} would indicate that the global minimum mass ratio is of little practical use and instead orbital instability onset is dependent on the mass of the primary component and that for low mass systems instability can occur with mass ratios as high as 0.22 to below 0.05. 

The number of identified contact binaries is ever increasing with new systems being continually added through various sky surveys. Large scale high resolution radial velocity observations are at present impractical thus limiting our search for potential merger candidates among systems demonstrating total eclipses and therefore suitable for light curve analysis. To this end \citet{2022JApA...43...94W} have introduced simplified techniques to identify potential extreme low mass ratio system from survey photometric data. The present study continues our programme of follow-up dedicated observations of potential merger candidates identified from such data. We report photometry and spectroscopic observations of six contact binaries identified as potential extreme low mass ratio systems from the ASAS-SN survey. All six are confirmed as being of extreme low mass ratio,  having a mass ratio less than 0.15. Three systems fall into the instability category based on theoretical considerations. It must be noted that the instability mass ratio is highly dependent on the mass of the primary and a 10\% change in the primary's mass can result in up to a 17\% change in the instability mass ratio \citep{2022MNRAS.512.1244C} so confirmation of the mass of the primary, in particular for three potentially unstable systems (A0842, A1037 and V565 Dra) through high resolution spectral observations would be desirable. 

Compared to other contact binaries the six systems reported here follow similar characteristics with significantly larger and brighter secondaries relative to main sequence counterparts. In addition, the secondary component in all cases is significantly denser than the primary component. Extreme low mass ratio contact binaries usually show photospheric signs of increased magnetic and chromospheric activity as a variation in the two maxima in the light curve due to the presence of star spots. The current sample of six did not show significant variation in maxima, however, non-photospheric markers such as increased high energy emissions, particularly in the far ultraviolet band, were present in all five of the systems observed by the GALEX mission. 

Recent progress in the rapid identification of low mass ratio systems from survey photometry and theoretical considerations for orbital stability has significantly increased the detection of potentially unstable system. As noted above, we have already reported 15 such systems and the current study adds a further three. The study also highlights that not all extreme low mass ratio systems will be unstable, three systems from the current study and some previous large surveys with over 10 systems in each study did not detect any unstable system even though they all reported extreme low mass ratio systems \citep{2023MNRAS.519.5760L, 2021MNRAS.502.2879G, 2022MNRAS.512.1244C, 2022AJ....164..202L}.

\begin{acknowledgements}

\noindent Acknowledgements.

\noindent Based on data acquired on the Western Sydney University, Penrith Observatory Telescope. We acknowledge the traditional custodians of the land on which the Observatory stands, the Dharug people, and pay our respects to elders past and present.\\

\noindent This research has made use of the SIMBAD database, operated at CDS, Strasbourg, France.\\

\noindent This work makes use of observations from the Las Cumbres Observatory global telescope network.\\

\noindent This publication makes use of VOSA, developed under the Spanish Virtual Observatory (https://svo.cab.inta-csic.es) project funded by MCIN/AEI/10.13039/501100011033/ through grant PID2020-112949GB-I00. VOSA has been partially updated by using funding from the European Union's Horizon 2020 Research and Innovation Programme, under Grant Agreement number 776403 (EXOPLANETS-A).\\

\noindent B. Arbutina acknowledges the funding provided by the Ministry of Science, Technological Development and Innovation of the Republic of Serbia through the contract  451-03-47/2023-01/200104.\\

\noindent During work on this paper, G. Djurašević and J. Petrović were financially supported by the Ministry of Science, Technological Development and Innovation of the Republic of Serbia through contract 451-03-47/2023-01/200002

\end{acknowledgements}


\bibliography{sample631}{}
\bibliographystyle{aasjournal}



\end{document}